\documentclass[%
mathleft,%
]{an}
\usepackage{graphicx}
\usepackage{times}
\begin{document}

% The following seven commands are intended for editorial usage and
% should be ignored by the author(s).
\Pagespan{1}{}% Document's page range. 
% If second parameter is left empty, the last page is computed
% automatically.
\Yearpublication{2011}%
\Yearsubmission{2011}%
\Month{1}%   
\Volume{999}%  
\Issue{92}% 
% \DOI{This.is/not.aDOI}% 

\title{A multiwavelength view of star-disk interaction in NGC 2264
\thanks{Based on data from the {\em Spitzer} and CoRoT missions. The CoRoT  space mission was developed and is operated by the French space
agency CNES, with participation of ESA's RSSD and Science Programmes, Austria,
Belgium, Brazil, Germany, and Spain.}}

\author{A.\,M.\, Cody\inst{1}\fnmsep\thanks{Corresponding author:
  \email{amc@ipac.caltech.edu}}
% Example for footnote, note the usage of the \texttt{fnmsep} command
% as separator between institute number and footnote mark}
\and  J.\,R. Stauffer\inst{1}
\and G. Micela\inst{2}
%\and M. Morales-Calder\'{o}n\inst{3}
%\and J. Bouvier\inst{4}
%\and S.\, P.\, H.\, Alencar\inst{5}
%\and E. Flaccomio\inst{2}
%\and K. Zwintz\inst{6}
%\and S. Aigrain\inst{7}
%\and L.\, M.\, Rebull\inst{1}
%\and L.\, A.\, Hillenbrand\inst{8}
%\and P.\, S.\, Teixeira\inst{6}
%\and D. Barrado\inst{3}
\and A. Baglin\inst{3}
\and the CSI 2264 Team\thanks{{\em http://csi2264.ipac.caltech.edu}}
}
\titlerunning{}
\authorrunning{A.\,M.\, Cody et al.}
\institute{Spitzer Science Center, California Institute of Technology, 1200 E California Blvd., Pasadena, CA 91125, USA
\and 
INAF-Osservatorio Astronomico di Palermo, Piazza del Parlamento 1, 90134 Palermo, Italy 
\and
LESIA, Université Pierre et Marie Curie, Université Denis Diderot, Observatoire de Paris, 92195 Meudon Cedex, France
%\and
%Centro de Astrobiología (INTA-CSIC), ESAC Campus, P.O. Box 78, E-28691 Villanueva de la Canada, Spain
%\and
%UJF-Grenoble 1/CNRS-INSU, Institut de Planétologie et d'Astrophysique de Grenoble (IPAG) UMR 5274, F-38041, Grenoble, 
%France
%\and
%Departamento de Física – ICEx – UFMG, Av. Antônio Carlos 6627, 30270-901 Belo Horizonte, MG, Brazil 
%\and
%University of Vienna, Institute for Astronomy, Türkenschanzstrasse 17, A-1180 Vienna, Austria
%\and
%Department of Physics, University of Oxford, Denys Wilkinson Building, Keble Road, Oxford OX1 3RH
%\and
%California Institute of Technology, MC 249-17, 200 E California Blvd., Pasadena, CA 91125, USA
}

\received{XXXX}
\accepted{XXXX}
\publonline{XXXX}


\abstract{
Variability is a signature property of cool young stars, particularly for 
those surrounded by disks. Traditional single-band time series display 
complex features associated with accretion, disk structure, and accompanying 
stellar activity, but these processes are challenging to model. To make 
progress in connecting observed time domain properties with the underlying 
physics of young stars and their disks, we have embarked on an unprecedented 
multiwavelength monitoring campaign: the Coordinated Synoptic Investigation 
of NGC 2264 ("CSI 2264"). Beginning in December 2011, CSI 
2264 has acquired 30 continuous days of mid-infrared time series from 
{\em Spitzer}, simultaneous optical monitoring from CoRoT and {\em MOST}, X-ray 
observations with Chandra, as well as complementary data from a number of 
ground-based telescopes. The extraordinary photometric precision, cadence, and 
time baseline of these observations enable detailed 
correlation of variability properties at different wavelengths, 
corresponding to locations from the stellar surface to the inner AU of 
the disk. We present the early results of the program, and discuss the
need for further modeling efforts into young stars and their disks.
}

\maketitle

\section{The promise of multiwavelength time series monitoring of young stars}

The canonical picture of a young accreting star (see Hartmann 1998) involves emission at a wide range of wavelengths, 
characterizing various regions from the central object to the outer reaches of its disk. Stellar flux comes primarily 
in the optical, with some contributions from magnetic spots and flares. Where the magnetosphere anchors to the 
surface, accretion material from the disk is thought to funnel along columns before colliding with a 
shocked region where ultraviolet radiation is produced. Further out, emission lines such as H$\alpha$ arise from ionized gas 
in the accretion flow. The disk is heated by the central star, and its innermost portions ($d<1$~AU) reradiate at 
near-infrared wavelengths. An inner wall may cast a shadow on the outer parts, which emit in the mid to far-infrared 
according to the lower dust temperatures found there. The measured flux at all of these wavelengths is further
dependent on the observer's aspect angle to the star/disk system as well as on the rotation periods of the emitting 
regions.

While this model involves a relatively static geometry, young stars and their disks constitute an incredibly dynamic 
environment, as borne out by variability studies. It has been known for decades (e.g., Joy 1945) that 
young stellar objects (YSOs) display prominent optical brightness fluctuations on timescales from days to years. 
Light curves contain not only regular sinusoidal patterns, but abrupt and unpredictable changes as 
well (e.g., Cody \& Hillenbrand 2010). More recently, it has become evident that many of these objects also
exhibit significant variations in the near- and mid-infrared, suggestive of changes in emission from the inner disk
itself (Morales-Calder\'{o}n et al.\ 2011; Rebull 2011).

A key question is how to connect variability with the physical configuration and processes relevant to YSOs. 
Initial attempts to correlate the optical and near-infrared time-domain properties of young stars with models (e.g., 
Herbst et al. 1994; Carpenter et al.\ 2001) have revealed photometric behavior that is consistent with variable accretion, 
hot and cool photospheric spots, or variable obscuration by circumstellar material. Yet with limited wavelength coverage 
or temporally sparse data, these scenarios could not be distinguished unambiguously. Further work on class II sources by 
Eiroa et al. (2002), Bary et al.\ (2009), and Espaillat et al. (2011) uncovered near-IR and mid-IR flux changes 
implicating disk thermal and structural changes on timescales from days to years.  It has been proposed that the more 
rapid variations reflect changes in the height of the inner disk wall (Hirose \& Turner 2011; Ke et al.\ 2012). Additional 
modeling efforts such as those by Dullemond et al.\ (2003), Flaherty \& Muzerolle (2010), and Romanova et al.\ (2011) have 
begun to offer descriptions of inner disk dynamics and star-disk interaction but require more extensive input from 
observations on varied timescales and wavelengths.

\begin{table} 
% \centering%%% 
\caption{Optical and infrared space-based observations} 
\label{tlab} 
\begin{tabular}{cccc}
\hline 
Telescope & \# of targets & Precision (mag)& Cadence (min)\\ 
\hline 
{\em Spitzer}/map & 1000 & 0.01--0.03 & 100 \\ 
{\em Spitzer}/stare & 540 & 0.001-0.01 & 0.1 or 1 \\ 
CoRoT & 500 & 0.0005--0.01 & 8.5 \\ 
MOST & 67 & 0.001-0.01 & 0.4 or 0.85 \\ 
\hline 
\end{tabular} 
\end{table}

\section{The Coordinated Synoptic Investigation of NGC 2264}
We have embarked on an unprecedented exploration of young star variability via high-precision, simultaneous optical 
and infrared time series monitoring of YSOs in NGC 2264. This few-Myr-old cluster contains some 2000 known members, 
many of which have disks (Rebull et al.\ 2002; Dahm \& Simon 2005). It was previously monitored in the optical (Lamm 
et al.\ 2004; Cieza \& Baliber 2007) as well as a ``short run'' (23 days) with the CoRoT satellite (Favata et 
al. 2010). Results from the latter program have contributed vitally to our understanding of the complexities of YSO 
variability at on timescales from minutes to weeks (e.g., Alencar et al.\ 2010, Zwintz et al.\ 2011).

Our campaign-- the Coordinated Synoptic Investigation of NGC 2264 (``CSI 2264'')-- combines the power of precision 
space-based photometry with the benefits of multiwavelength monitoring. The program commenced in early December 2011, 
with roughly 30 continuous days of mid-infrared photometry from {\em Spitzer}/IRAC and 40 continuous days of optical 
monitoring with CoRoT, targeting the central degree of the cluster. Four $\sim$1-day blocks were dedicated to 
monitoring of two 5.2\arcmin\ regions near the cluster center with {\em Spitzer's} high precision ($<$1\%) staring 
mode, whereas the remaining targets were visited in mapping mode (1--3\% precision). Complementing these 
observations were 40 days of high-cadence, high-precision time series of 67 of the brightest cluster members with the 
Microvariability and Oscillations of STars telescope ({\em MOST}; Walker et al.\ 2003), as well as 350~ks ($\sim$4 
days) of {\em Chandra}/ACIS X-ray monitoring simultaneous with the {\em Spitzer} staring observations. In addition, 
synoptic ground-based optical and near-infrared data in the $U$ through $K$ bands was acquired simultaneously with a 
number of instruments, including $R\sim 17,000$ spectra from the VLT/Flames multi-object spectrograph as well as 
optical photometry from the USNO 40-inch telescope ($I$ band), and the Canada-France-Hawaii Telescope MegaCam ($U$, 
$R$ band). The bulk of this auxiliary monitoring continued through February 2012.

The space-borne instruments involved in CSI~2264 have a history of providing exquisite precision photometric time 
series at minute cadences (see Table 1). Since NGC 2264 is the only young open cluster available for simultaneous 
monitoring by {\em Spitzer} and CoRoT, we expect the combined dataset to provide insights into the dynamic 
environment of young stars and their disks for years to come. The project is still in its early stages, and we 
present here some initial insights gleaned from the combination of optical and mid-infrared time series.

%\subsection{{\em Spitzer} observations}

%precision
%magnitudes
%staring (correction)
%mapping
%bands

%\subsection{CoRoT observations}

\section{Initial results from CSI 2264}

\begin{figure*}
\includegraphics[width=\linewidth,height=50mm]{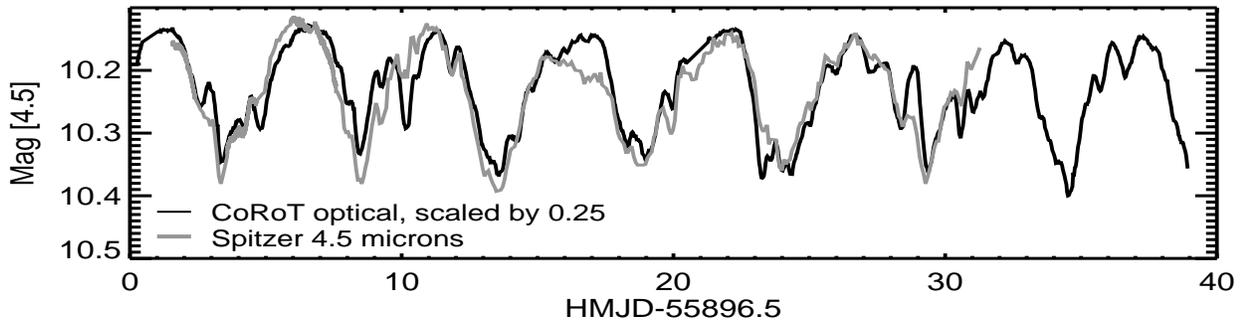}
\caption{The optical light curve of NGC 2264 member V354~Mon from CoRoT (black curve), with flux scaled by a factor  
of 0.25 and shifted so as to have the same median of the {\em Spitzer}/IRAC 4.5~$\mu$m data. The two light curves   
display variability that is well but not perfectly correlated on timescales of $\sim$5 days. The scale factor is
significantly different from the ratio of $A_{4.5}/A_V$ expected from an interstellar extinction law ($\sim$0.05).}
\label{label1}
\end{figure*}

\begin{figure*}
\includegraphics[width=\linewidth,height=50mm]{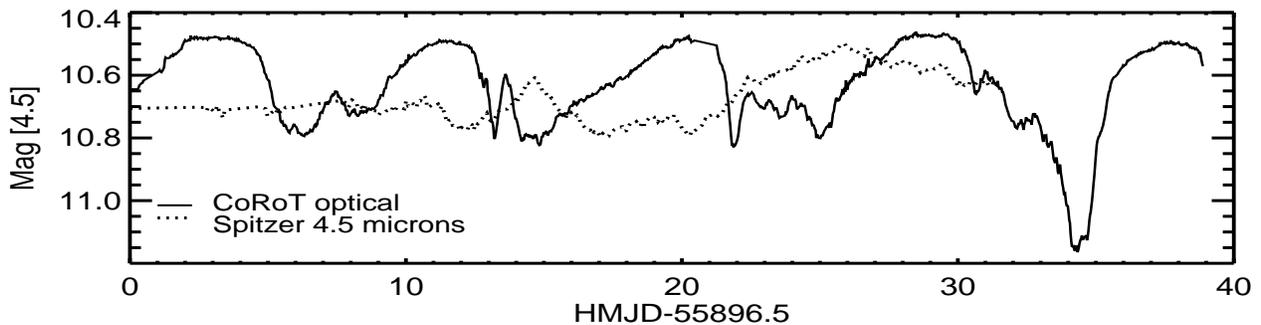}
\caption{An example of uncorrelated variability in optical and mid-infrared. The CoRoT data is median shifted to match
the {\em Spitzer} zeropoint. The contrast between this behavior and the high degree of correlation in Fig. 1 might be 
accounted for by a different aspect angle to the disk and/or a larger fraction of the 4.5~$\mu$m luminosity coming from the 
inner disk, as opposed to the central star.}
\label{label1}
\end{figure*}

%We highlight here some of the pronounced variability patterns uncovered thus far in the combined CoRoT/{\em Spitzer} 
%dataset.
\subsection{Periodic and semi-periodic variability}

Periodic variability in YSOs is usually attributed to rotational modulation of the light curve by magnetic spots on the 
stellar surface. Many of our light curves display sinusoidal variations in both bands, and this is particularly 
characteristic of the type III (i.e., no infrared excess) objects that are presumed to lack disks. Typically, the IRAC 
amplitudes are lower than those in the optical CoRoT light curve, but the variations become more comparable toward 
later spectral type. A handful of objects have more than one significant period in their light curve, suggesting that 
spots are present at multiple latitudes, or that we are observing a binary system involving two active stars with different 
rotation periods. A number of eclipsing binaries also lie in the NGC 2264 field, and follow-up monitoring is expected 
to provide precise stellar parameters for a number of them, along with new benchmarks for comparison with models at 
young ages. At the bright end of the sample, asteroseismology of cluster $\delta$ Scuti stars will continue to unveil 
the properties of their interiors (e.g., Zwintz et al.\ 2011).

A subset of objects display semi-periodic variability in both bands that is not necessarily consistent with the spot 
modulation scenario. Much of this is large amplitude ($>$0.1 magnitudes) and involves a combination of repeating features 
as well as smaller amplitude deviations which appear and disappear on $\sim$1 day timescales. We suspect that some of this 
behavior may be explained by the periodic passage of obscuring disk material by the face of the central star. This was the 
idea put forth to explain the ``AA Tau'' phenomenon, which was highlighted in a previous CoRoT short run dataset on 
NGC~2264 by Alencar et al.\ (2010). A prominent example of AA~Tau type behavior occurs in NGC~2264 member V354~Mon, which 
displays $\sim$0.25 magnitude fluctuations in the optical and $\sim$0.06 magnitude fluctuations in the infrared. The 
variations in the two bands mirror each other quite well when those from the 4.5~$\mu$m band are scaled up by a factor of 
4.0, as seen in Fig.\ 1. Supposing that the brightness variations are explained solely by changing dust extinction in 
the line of sight to the star, we arrive at a reddening law, $A_{4.5}/A_V$, that is approximately five times larger 
than the standard interstellar predictions (e.g., Indebetouw et al. 2005). Taking into account the 4.5~$\mu$m flux 
from the disk itself would only raise this value further. Therefore if these semi-periodic changes are indeed due to 
obscuration by disk material, then we can infer that its dust properties are significantly different from those of ISM grains.

In contrast, other cases of semi-periodic variability involve relatively colorless brightness fluctuations. The opacity 
of obscuring material may vary from star to star and play a role in determining the color trends in their light curves.

\subsection{Aperiodic variability}

The majority of disk-bearing stars in NGC~2264 display significant variability at both optical and mid-infrared 
wavelengths that contains aperiodic features. In most cases, the behavior in the two bands appears at least somewhat 
correlated. This is particularly evident for YSOs whose optical light curves display deep ($>$0.1~mag), semi-periodic 
fading events consistent with the AA~Tau phenomenon. However, we also identify more extreme cases of 
variability, in which the brightness in both the optical and mid-infrared varies by more than 20\%, but fluctuations 
appear completely uncorrelated at these two different bands. We present an example in Fig.\ 2. The largest 
amplitude infrared variability in these objects occurs on longer timescales (5 days or more) than that seen 
in the semi-periodic objects. Most of the optical light curves, on the other hand, contain high amplitude dip-like or 
undulating features on 1--5 day time scales.

A further subclass of variables exhibits large-amplitude variation ($>$0.2 magnitudes) in the IRAC bands but relatively 
little variation as seen by CoRoT.  An example is shown in Fig.\ 3.  The appearance of the mid-infrared light curves 
in this class varies from large-amplitude excursions on 3--5 day timescales to more smoothly changing brightness on 
longer timescales. In cases for which fluctuations take place over several days, variability behavior evolves faster 
than the dynamical timescale of the inner disk. This mid-infrared flux variation may involve a combination of 
structural, dynamical, and thermal variations in the disk but currently lacks more detailed explanation.

Many of the cluster members monitored are accreting, based on strong H$\alpha$ emission and $U$-band excesses. The 
accretion process likely proceeds in bursts, and changes in accretion luminosity may be another source of stochastic 
variability in YSOs. We tentatively identify a number of cases for which the optical brightness undergoes abrupt 
brightening events consistent with accretion changes. Flux outbursts typically last several days, with lower level 
structure on shorter timescales. Nearly all of these objects display significant UV excesses, as inferred from our 
CFHT/Megacam $U$-band dataset. Figure 4 illustrates an example of behavior that we attribute to accretion. 
Where IRAC photometry is available and unsaturated, the mid-infrared flux typically displays similar abrupt increases, 
probably reflective of disk heating following an increase in accretion luminosity. 

\begin{figure*}
\includegraphics[width=\linewidth,height=48mm]{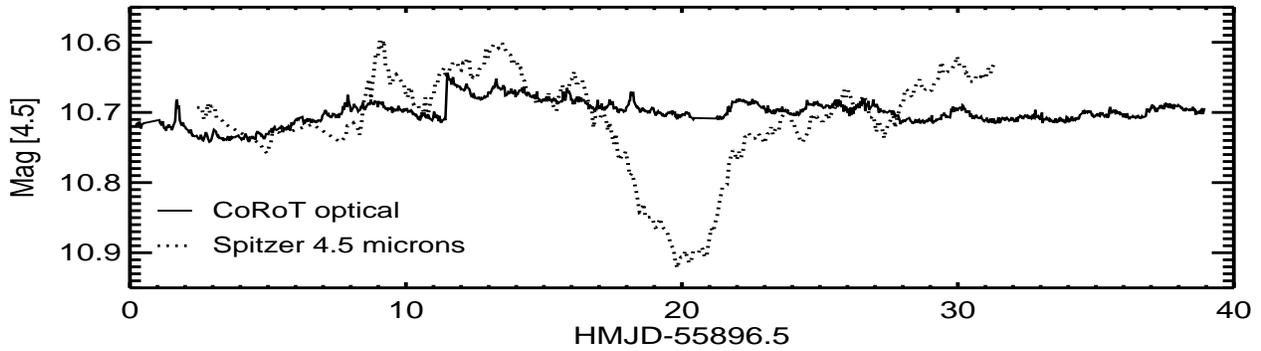}
\caption{An example in which mid-infrared variability appears with large amplitude, whereas optical brightness
fluctuations occur at a much lower level.}
\label{label1}
\end{figure*} 

\begin{figure*}
\includegraphics[width=\linewidth,height=48mm]{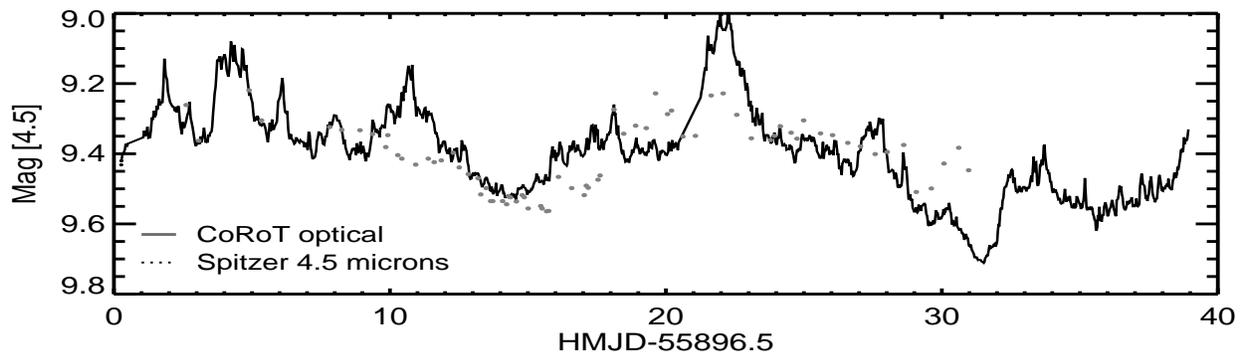}
\caption{NGC 2264 member for which abrupt increases in optical brightness are suggestive of accretion bursts.}
\label{label1}
\end{figure*}

\section{Summary}
The CSI 2264 project offers great potential to unlock the multiwavelength time-domain properties of YSOs across a 
range of stellar and disk properties. Further investigation of this high-precision, high-cadence dataset is 
expected to shed light on the physical mechanisms of variability and potentially reveal the properties of 
otherwise inaccessible inner disk regions. The eventual inclusion of high-resolution spectroscopic data for a 
subset of targets, as well as photometry at other bands from the X-ray through near-infrared, will enhance these 
efforts. Ultimately, reduced data from the campaign will be made publicly available from a website hosted at the 
NASA/Caltech Infrared Processing and Analysis Center.

\acknowledgements
This work is based on observations made with the CoRoT satellite and the Spitzer Space
Telescope. {\em Spitzer} is operated by the Jet Propulsion Laboratory, California Institute of
Technology under a contract with NASA. Support for this work was provided by NASA
through an award issued by JPL/Caltech.

% Use this code if you wish to generate your bibliography with BibTeX;
% please replace first the string "an-demo" below with the name(s) of
% the BibTeX data base(s) you want to use.
% The resulting bibliography-output (the contents of the .bbl file)
% must be pasted into this file before submission.
% 
% \bibliographystyle{an}
% \bibliography{an-demo}
% 
% Replace the following example bibliography with your references
% before submission:

\end{document}